\documentclass[aip,pop,reprint]{revtex4-1}
 \usepackage{dcolumn}
 \usepackage{bm}
 \usepackage{amsmath}

\newcommand{\vc}{\mathbf}

\newcommand{\la}{\ensuremath{\langle}}
\newcommand{\ra}{\ensuremath{\rangle}}

\newcommand{\qm}{\textrm{qm}}
\newcommand{\cl}{\textrm{cl}}
\newcommand{\D}{\textrm{D}}

\usepackage{graphics}
\usepackage{color}

\begin{document}


\title{A generalized Coulomb logarithm for strongly coupled plasmas}


\author{Scott D. Baalrud}

\affiliation{Center for Integrated Computation and Analysis of Reconnection and Turbulence, University of New Hampshire, Durham, New Hampshire 03824}

\date{\today}

\begin{abstract}

Transport coefficients for Coulomb collision processes, e.g., friction, energy exchange, and resistivity, are calculated for Debye-screened (Yukawa) plasmas including correlation effects within the binary collision approximation. A generalized Coulomb logarithm, $\Xi(\Lambda)$, is required, which asymptotes to $\ln \Lambda$ in the weakly coupled limit, but is significantly modified when $\Lambda \lesssim 10^2$ due to large angle collisions. In the regime $10^{-2} \lesssim \Lambda \lesssim 10^2$, $\Xi$ can be substantially larger for attractive collisions than repulsive. 

\end{abstract}



\maketitle

Plasmas in several modern research areas, including inertial confinement fusion (ICF),\cite{atze:04} antimatter plasmas,\cite{andr:10} ultracold plasmas,\cite{kill:07} and dusty plasmas,\cite{khra:09} can enter regimes in which the central tenet of conventional plasma kinetic theory is violated. Conventional theories~\cite{land:36,rose:57,lena:60} are based upon a weak coupling approximation, often referred to as ``the plasma approximation,'' which assumes $\ln \Lambda \gg 1$, where $\Lambda \simeq 12 \pi n \lambda_\D^3 \sim T^{3/2}/(Z_i \sqrt{n})$ is the plasma parameter, and $\lambda_\D = (\sum_s \lambda_{\D s}^{-2})^{-1/2}$ is the Debye length ($\lambda_{\D s}^2 = T_s/4\pi q_s^2 n_s$). Most plasmas in space and the laboratory are weakly coupled, $\ln \Lambda \gtrsim 10$, but moderate, $2 \lesssim \ln \Lambda \lesssim 10$, and strong, $\ln \Lambda \lesssim 2$, coupling regimes can be reached if the plasma is extremely dense (ICF), cold (antimatter and ultracold), or highly charged (dusty). Understanding how Coulomb collisions affect transport properties, such as friction, temperature equilibration, and electrical conductivity, is essential in order to accurately predict the behavior of these plasmas. 

A generalized method of calculating transport coefficients for fluid equations, which relaxes the conventional weak coupling approximation, is provided. The weak coupling approximation is utilized twice in conventional theories, and both instances are avoided here. The first instance is that conventional theories are based on a small scattering angle expansion, or a hierarchy expansion (BBGKY), which orders terms according to $\ln \Lambda \gg 1$. The second instance is that a logarithmically divergent integral arises as a result of using the bare Coulomb potential for individual particles. Debye screening is introduced, and the integral kept finite, by imposing an ad-hoc cutoff for the maximum impact parameter at the Debye length. This cutoff is only justified when $\ln \Lambda \gg 1$. Here, we avoid the small scattering angle expansion by exploiting symmetries that arise in velocity-space moments of the full Boltzmann collision integral. The theory self-consistently accounts for Debye screening, and thus avoids divergent integrals, by using the screened Coulomb potential. 

The screened Coulomb potential is common, but not universal, in the plasmas mentioned above. For example, collective interactions can cause more complex behavior than Debye screening, such as wakes surrounding fast-flowing particles.\cite{milo:10} These effects are especially important for fast particle stopping power in ICF. However, it is a valid approximation for many processes of interest, regardless of correlation strength. 

We find that a generalized Coulomb logarithm, $\Xi$, is required. This asymptotes to $\Xi \simeq \ln \Lambda$ in the weakly coupled limit, but is modified significantly by correlation effects when $\Lambda \lesssim 10^2$. In the regime, $10^{-2} \lesssim \Lambda \lesssim 10^2$, $\Xi$ can be substantially larger for attractive collisions than repulsive. The present work focuses on species with flowing Maxwellian distributions. However, the underlying approach does not depend on the functional form of the distributions and can be applied more generally. One such example is Spitzer's approach,\cite{spit:53} which considers deviations from Maxwellian as higher-order perturbations. 

The theory is based on the Boltzmann collision operator, which requires that the plasma be sufficiently dilute to justify the binary collision approximation. Effects of three, and more, particle collisions, as well as phase transitions, become important at sufficiently strong coupling. Nevertheless, this approach is valid for coupling parameters ($\Lambda^{-1}$) orders of magnitude larger than conventional theory can describe. To demonstrate this, we compare the calculated temperature relaxation rate between electrons and ions for ICF relevant parameters with recent molecular dynamics (MD) simulations.\cite{dimo:08} Our theory agrees with the simulation results for all values of $\Lambda$ that were simulated ($10^{-1} \lesssim \Lambda \lesssim 10^{3}$), whereas conventional theories widely disagree with them for $\Lambda \lesssim 10$.  Fluid codes used to simulate strongly coupled plasmas require transport coefficients that can extend beyond the weakly coupled Landau-Spitzer theory.\cite{land:36,spit:53} Current fluid codes disagree with MD simulations of thermal equilibration rates, hot spot formation, laser absorption, and neutron yield in correlated regimes; processes essential to accurate modeling of ICF implosions.\cite{bene:09,xu:11,hans:11}

The Boltzmann kinetic equation describing the evolution of the distribution function of a species $s$ is $\partial_t f_s + \vc{v} \cdot \nabla f_s + \vc{a} \cdot \nabla_{\vc{v}} f_s = \sum_{s^\prime} C_\textrm{B}(f_s, f_{s^\prime})$,
in which
\begin{equation}
C_\textrm{B}(f_s, f_{s^\prime}) = \int_{\vc{v}^\prime} \int_{\Omega} d^3v^\prime d\Omega\, \sigma\, u\, (\hat{f}_s \hat{f}^\prime_{s^\prime} - f_s f_{s^\prime}^\prime)  \label{eq:cb}
\end{equation}
is the operator describing collisions between species $s$ and $s^\prime$. Here $\hat{f}_s = f_s(\hat{\vc{v}})$, $\hat{f}_{s^\prime}^\prime = f_{s^\prime} (\hat{\vc{v}}^\prime)$, $f_s = f_s(\vc{v})$, $f_{s^\prime}^\prime = f_{s^\prime}(\vc{v}^\prime)$, in which $(\vc{v}, \vc{v}^\prime)$ are the initial particle velocities, and $(\hat{\vc{v}}, \hat{\vc{v}}^\prime) = (\vc{v} + \Delta \vc{v}, \vc{v}^\prime + \Delta \vc{v}^\prime)$ are the particle velocities after a collision. The differential scattering cross section is denoted $\sigma$, the solid angle $d\Omega = d\phi d\theta \sin \theta$, the relative velocity $\vc{u} \equiv \vc{v} - \vc{v}^\prime$, and $u \equiv |\vc{u}|$.  Fluid equations of motion are derived from velocity-space moments of the plasma kinetic equation.\cite{brag:65} Here we are interested in the effects of Coulomb collisions, which enter through the friction force density $\vc{R}^{s-s^\prime} = \int d^3v\, m_s \vc{v} C(f_s, f_{s^\prime})$, energy exchange density $Q^{s-s^\prime} = \int d^3v\, \frac{1}{2} m_s (\vc{v} - \vc{V}_s)^2 C(f_s,f_{s^\prime})$, and so on for higher-order moments. The reference resistivity, $\eta_o$, can be obtained from the electron-ion friction force density, $\vc{R}^{e-i} \approx \eta_o en_e \vc{J}$, where $\vc{J} = en_e (\vc{V}_i - \vc{V}_e)$ is the current density. 

The collision operator used to evaluate $\vc{R}^{s-s^\prime}$ and $Q^{s-s^\prime}$ in weakly coupled plasma theory~\cite{land:36,rose:57} can be obtained from Eq.~(\ref{eq:cb}) as follows. After applying a small scattering angle expansion $(\vc{v}, \vc{v}^\prime) \gg (\Delta \vc{v}, \Delta \vc{v}^\prime)$, integrals of the form
\begin{equation}
\lbrace \Delta \vc{u} \rbrace = \int d\Omega\, \sigma\, u\, \Delta \vc{u} , \label{eq:du}
\end{equation} 
as well as $\lbrace \Delta \vc{u} \ldots \Delta \vc{u} \rbrace$ arise,\cite{rose:57} in which $\Delta \vc{u} = u[\sin \theta\, \cos \phi\, \hat{x} + \sin \theta\, \sin \phi\, \hat{y} - 2 \sin^2 (\theta/2) \hat{u}]$ is the rotation of the relative velocity vector after scattering through angle $\theta$. Typically, the Rutherford scattering formula $\sigma = q_s^2 q_{s^\prime}^2/[4 m_{ss^\prime}^2 u^4 \sin^4(\theta/2)]$ is used in~Eq.~(\ref{eq:du}) and Debye screening is introduced by assuming the maximum impact parameter is the Debye length. Doing so yields $\lbrace \Delta \vc{u} \rbrace = -4\pi q_s^2 q_{s^\prime}^2 \ln (\lambda_\D/b_{\min})/(m_{ss^\prime}^2 u^2)$, where $b_{\min}^{\cl} = q_s q_{s^\prime}/(m_{ss^\prime} u^2)$. If $b_{\min}^{\cl}$ is smaller than the de Broglie wavelength, $b_{\min}^{\qm} = \hbar/(2m_{ss^\prime} u)$ is used. 

Using the above approximations, and keeping only terms of $\mathcal{O}[\ln (\lambda_\D/b_{\min})]$, the collision operator can be written in the Landau form.\cite{land:36,lifs:81} Taking the distribution functions to be flowing Maxwellians, $f_s(\vc{v}) = n_s \exp[-(\vc{v} - \vc{V}_s)^2/v_{Ts}^2]/\pi^{3/2} v_{Ts}^3$, the friction force density and energy exchange density are found to be
\begin{equation}
\vc{R}^{s-s^\prime} = - n_s m_s \nu_{s-s^\prime} (\vc{V}_s - \vc{V}_{s^\prime}) \label{eq:rssp}
\end{equation}
and
\begin{equation}
Q^{s-s^\prime} = -3 \frac{m_{ss^\prime}}{m_{s^\prime}} n_s \nu_{s-s^\prime} (T_s - T_{s^\prime}) , \label{eq:qssp}
\end{equation}
to lowest order in $\ln \Lambda \gg 1$, and $|\vc{V}_s - \vc{V}_{s^\prime}|/\bar{v}_T\ll 1$. The characteristic collision rate is 
\begin{equation}
\nu_{s-s^\prime} = \frac{16 \sqrt{\pi} q_s^2 q_{s^\prime}^2 n_{s^\prime}}{3 m_s m_{ss^\prime} \bar{v}_T^3} \Xi \label{eq:nussp}
\end{equation}
where $\bar{v}_T^2 = v_{Ts}^2 + v_{Ts^\prime}^2$. In this weakly coupled limit, 
\begin{equation}
\Xi = \Xi_{\textrm{wc}} = \ln \Lambda  , \label{eq:xiwc}
\end{equation}
in which $\Lambda$ is the minimum of $\Lambda_{\cl} = m_{ss^\prime} \lambda_\D \bar{v}_T^2 /|q_s q_{s^\prime}|$ or $\Lambda_{\qm} = 2m_{ss^\prime} \lambda_\D \bar{v}_T/\hbar$.

Previous authors have improved the conventional theory. Lenard and Balescu~\cite{lena:60} accounted for the dielectric nature of a plasma with a method that self-consistently obtains $b_{\max} = \lambda_\D$ in the adiabatic limit, but misses $b_{\min}$ and still diverges. Aono later showed that the Lenard-Balescu and Landau approaches can be combined to resolve the logarithmic divergence issue, and others have obtained similar results with various methods.\cite{aono:68,libo:59,lee:84,lifs:81,brow:05} These theories also obtain order unity corrections that give $\Xi = \ln(C\Lambda)$, where $C\approx 0.765$. Although these self-consistently derive $\ln \Lambda$, they are limited to weak coupling because they rely on a small angle scattering expansion. Li and Petrasso~\cite{li:93a} attempted to describe moderately coupled plasmas by extending the expansion procedure to third order; again obtaining an order unity correction. However, although the first two terms of this expansion are the only ones with $\mathcal{O}(\ln \Lambda)$ contributions, every other term in the expansion has an order unity contribution. One must both abandon the small angle scattering expansion, and obtain a convergent collision integral, to include correlation effects.

The symmetry properties of the scattering process that we use are the same as those used to derive the Boltzmann collision operator itself: (i) the total differential velocity-space volume element is invariant, $d^3\hat{v} d^3 \hat{v}^\prime = d^3v d^3v^\prime$, (ii) the magnitude of the relative velocity vector is invariant, $\hat{u} = u$, and (iii) the ``inverse'' collision satisfies $\hat{\sigma} (\vc{v}, \vc{v}^\prime \rightarrow \hat{\vc{v}}, \hat{\vc{v}}^\prime) = \hat{\sigma}(\hat{\vc{v}}, \hat{\vc{v}}^\prime \rightarrow \vc{v}, \vc{v}^\prime)$.\cite{reif:65} Here, $\hat{\sigma}(\vc{v}, \vc{v}^\prime \rightarrow \hat{\vc{v}}, \hat{\vc{v}}^\prime) d^3\hat{v} d^3v^\prime$, is the scattering probability, which is related to the differential scattering cross section, $\sigma$, by $\sigma d \Omega = \int d^3\hat{v} d^3\hat{v}^\prime\, \hat{\sigma}(\vc{v}, \vc{v}^\prime \rightarrow \hat{\vc{v}}, \hat{\vc{v}}^\prime)$. 

Transport coefficients can be determined from velocity moments of the form 
\begin{equation}
\la \chi \ra^{s-s^\prime} = \int d^3v\, \chi_s (\vc{v}) C_\textrm{B} (f_s, f_{s^\prime})  . \label{eq:meanchi}
\end{equation}
Here we will be interested in $\chi_s = (m_s, m_s \vc{v}, m_s v^2)$, which are related to density, momentum, and energy conservation. In terms of $\hat{\sigma}$, Eq.~(\ref{eq:meanchi}) is $\la \chi \ra^{s-s^\prime} = \int d^3v d^3v^\prime d^3\hat{v} d^3\hat{v}^\prime\, \hat{\sigma}(\vc{v}, \vc{v}^\prime \rightarrow \hat{\vc{v}}, \hat{\vc{v}}^\prime)\, u\, \chi_s(\vc{v}) [\hat{f}_s(\hat{\vc{v}}) \hat{f}_{s^\prime} (\hat{\vc{v}}^\prime) - f_s (\vc{v}) f_{s^\prime}(\vc{v}^\prime)]$. Making the interchange $(\hat{\vc{v}}, \hat{\vc{v}}^\prime) \leftrightarrow (\vc{v}, \vc{v}^\prime)$, and applying properties (i)--(iii) above, $\int d^3v d^3v^\prime d^3\hat{v} d^3\hat{v}^\prime\, \hat{\sigma}(\vc{v}, \vc{v}^\prime \rightarrow \hat{\vc{v}}, \hat{\vc{v}}^\prime)\, u \chi_s(\vc{v}) \hat{f}_s (\hat{\vc{v}}) \hat{f}_{s^\prime}(\hat{\vc{v}}^\prime) = \int d^3v d^3v^\prime d^3\hat{v} d^3\hat{v}^\prime\, \hat{\sigma}(\vc{v}, \vc{v}^\prime \rightarrow \hat{\vc{v}}, \hat{\vc{v}}^\prime)\, u \chi_s(\hat{\vc{v}}) f_s (\vc{v}) f_{s^\prime}(\vc{v}^\prime)$. Thus, Eq.~(\ref{eq:meanchi}) can be written in terms of the distribution functions before scattering 
\begin{equation}
\la \chi \ra^{s-s^\prime} = \int d^3v d^3v^\prime\, \lbrace \Delta \chi_s \rbrace f_s (\vc{v}) f_{s^\prime}(\vc{v}^\prime)
\end{equation}
in which $\lbrace \Delta \chi_s \rbrace = \int d\Omega\, \sigma\, u \Delta \chi_s$, and $\Delta \chi_s = \chi_s(\hat{\vc{v}}) - \chi_s(\vc{v})$. 

The friction force density is simply $\vc{R}^{s-s^\prime} = \la m_s \vc{v} \ra^{s-s^\prime}$. Conservation of momentum, $m_s \vc{v} + m_{s^\prime} \vc{v}^\prime = m_s \hat{\vc{v}} + m_{s^\prime} \hat{\vc{v}}^\prime$, implies $m_s\Delta \vc{v} = m_{ss^\prime} \Delta \vc{u}$, so 
\begin{equation}
\vc{R}^{s-s^\prime} = m_{ss^\prime} \int d^3u\, \lbrace \Delta \vc{u} \rbrace \int d^3v^\prime\, f_{s} (\vc{u} + \vc{v}^\prime) f_{s^\prime} (\vc{v}^\prime).  \label{eq:r2}
\end{equation}
The energy exchange density is $Q^{s-s^\prime} = \la \frac{1}{2} m_s v^2 \ra - \vc{V}_s \cdot \vc{R}^{s-s^\prime}$. Conservation of energy, $m_s v^2 + m_{s^\prime}v^{\prime 2} = m_s \hat{v}^2 = m_{s^\prime} \hat{v}^{\prime 2}$, implies $\Delta \vc{u} \cdot \Delta \vc{u} = -2 \vc{u} \cdot \Delta \vc{u}$, and $m_s \Delta v^2 = m_{ss^\prime} (\vc{v}^\prime + m_{ss^\prime} \vc{u}/m_{s^\prime}) \cdot \Delta \vc{u}$, so
\begin{equation}
Q^{s-s^\prime} = m_{ss^\prime} \int d^3u\, \lbrace \Delta \vc{u} \rbrace \cdot \vc{I}_\vc{u} , \label{eq:q2}
\end{equation}
where $\vc{I}_{\vc{u}} = \int d^3v^\prime ( \vc{v}^\prime - \vc{V}_s + m_{ss^\prime} \vc{u}/m_s) f_{s} (\vc{u} + \vc{v}^\prime)  f_{s^\prime} (\vc{v}^\prime)$.

Assuming the force between scattering particles is central and conservative, $\sigma = \sigma(|\vc{u}|, \theta)$, so $\lbrace \Delta \vc{u} \rbrace = -4\pi u\vc{u} \int_0^\pi d\theta\, \sin \theta\, \sin^2(\theta/2)\, \sigma(u,\theta)$. Taking $f_s$ and $f_{s^\prime}$ to be flowing Maxwellians, and assuming $|\vc{V}_s - \vc{V}_{s^\prime}|/\bar{v}_T \ll 1$, Eqs.~(\ref{eq:r2}) and (\ref{eq:q2}) can then be written in the identical form of Eqs.~(\ref{eq:rssp}) and (\ref{eq:qssp}), but with the generalized Coulomb logarithm
\begin{equation}
\Xi = \frac{1}{2} \int_0^\infty d\xi\, e^{-\xi^2}\xi^5 \frac{\sigma_s(\xi, \Lambda)}{\sigma_o} , \label{eq:xigen}
\end{equation}
replacing Eq.~(\ref{eq:xiwc}) for the collision frequency expression in Eq.~(\ref{eq:nussp}). In Eq.~(\ref{eq:xigen}), 
\begin{equation}
\sigma_s \equiv 4 \pi \int_0^\pi d\theta\, \sin \theta\, \sin^2(\theta/2)\, \sigma
\end{equation}
is the momentum-transfer cross section, $\sigma_o = \pi \lambda_\D^2/\Lambda_{\cl}^2$ is a reference value for the differential scattering cross section, and $\xi =u/\bar{v}_T$.

If the fluid transport timescale is long compared to the inverse plasma frequency $(\nu_{s-s^\prime}/\omega_{ps} \ll 1$), the polarization response of species $s$ is adiabatic and typically satisfies the Boltzmann relation $n_s = n_o \exp(-q_s \phi/T_s)$. As long as $q_s \phi/T_s \ll 1$, Poisson's equation $\nabla^2 \phi = -4\pi \rho_q$ shows that the potential around a test charge [$\rho_q = q_t \delta(\vc{x} - \vc{x}_t) + \rho_{\textrm{pol}}]$ in a plasma $\rho_{\textrm{pol}} = \sum_s q_s n_s$ has the Debye screened (Yukawa) form 
\begin{equation}
\phi_t = \frac{q_t}{r} e^{-r/\lambda_\D} , \label{eq:yukawa}
\end{equation}
where $r = |\vc{x} - \vc{x}_t|$ and $\lambda_{\D} = (\sum_s \lambda_{\D s}^{-2})^{-1/2}$. 

The Debye screened potential thus requires both adiabaticity ($\nu_{s-s^\prime}/\omega_{ps} \ll 1$) and a weak interaction potential ($q_s \phi/T_s \ll 1$). These conditions are often satisfied in both weakly and strongly coupled plasmas. For electron-ion collisions, Eq.~(\ref{eq:nussp}) shows $\nu_{e-i}/\omega_{pe} \sim \Xi/\Lambda$. Here, $\Lambda \sim n \lambda_\D^3$ has been used. For weak coupling ($\Lambda \gg 1$), $\nu_{e-i}/\omega_{pe} \sim \ln \Lambda/\Lambda \ll 1$. We will find that for strong coupling ($\Lambda \ll 1$), $\nu_{e-i}/\omega_{pe} \sim \Lambda \ln^2 \Lambda \ll 1$ [see Eq.~(\ref{eq:xicl_b})]. Thus, adiabaticity can be satisfied in both limits. For Eq.~(\ref{eq:yukawa}), $|q\phi/T| \sim (b_{\min}/r) \exp(-r/\lambda_\D)$. At the mean separation distance $r\sim n^{-1/3}$, $|q\phi/T| \sim \Lambda^{-2/3} \exp(-\Lambda^{-1/3})$, which is also small for both weak and strong correlation. The ratio of mean interaction distance ($n^{-1/3}$) to collision length ($\lambda_{e-i} \sim \nu_{e-i}/v_{Te}$) is also small in both regimes: $r/\lambda_{e-i} \sim \Xi/\Lambda^{4/3} \ll 1$.


\textit{Quantum regime.}--An analytic solution of Eq.~(\ref{eq:xigen}) can be obtained in the quantum regime $(\Lambda_{\qm} < \Lambda_{\cl})$. Using the Yukawa potential from Eq.~(\ref{eq:yukawa}) in the first Born approximation yields
\begin{equation}
\sigma_{\qm} = \frac{\sigma_o}{4\pi \xi^4} \frac{1}{[\Lambda_{\qm}^{-2} \xi^{-2} + \sin^2(\theta/2)]^2} . \label{eq:sigqm}
\end{equation}
The momentum scattering cross section for Eq.~(\ref{eq:sigqm}) is 
\begin{equation}
\sigma_s^{\qm} = \frac{2 \sigma_o}{\xi^4} \biggl[ \ln (\Lambda_{\qm}^2 \xi^2 + 1) - \frac{\Lambda_{\qm}^2 \xi^2}{1 + \Lambda_{\qm}^2 \xi^2} \biggr] . \label{eq:sigsqm}
\end{equation}
The effective Coulomb logarithm, $\Xi$, can be computed by putting Eq.~(\ref{eq:sigsqm}) into Eq.~(\ref{eq:xigen}), which yields
\begin{equation}
\Xi_{\qm} = \frac{1}{2} e^{\Lambda_{\qm}^{-2}} E_1(\Lambda_{\qm}^{-2}) (1 + \Lambda_{\qm}^{-2}) - \frac{1}{2}  , \label{eq:xiqm}
\end{equation}
where $E_1(x) = \int_{x}^\infty dt\, e^{-t}/t$ is the exponential integral. For $\Lambda_{\qm} \gg 1$, $\Xi_{\qm} = \ln \Lambda_{\qm} - (\gamma +1)/2 + \mathcal{O}(\Lambda_{\qm}^{-2})$, where $\gamma \simeq 0.577$ is Euler's constant.

\textit{Classical regime.}--For classical plasmas ($\Lambda_{\cl} < \Lambda_{\qm}$), it is convenient to write  Eq.~(\ref{eq:xigen}) in terms of the impact parameter $b$. Using $d\theta \sin \theta\, \sigma = b db$, and $\theta = \pi - 2 \Theta$, where 
\begin{equation}
\Theta =  b \int_{r_o}^\infty \frac{dr}{r^2 \sqrt{1 - U_{\textrm{eff}}(r,b)}} \label{eq:tcl}
\end{equation} 
and $r_o(b)$ solves $U_\textrm{eff} = 1$, the momentum-transfer cross section is 
\begin{equation}
\sigma_s = 4 \pi \int_0^\infty db\, b\, \cos^2 \Theta .
\end{equation}
For the Yukawa potential of Eq.~(\ref{eq:yukawa}), 
\begin{equation}
U_{\textrm{eff}} = \pm \frac{2}{\Lambda_{\cl} \xi^2} \frac{\lambda_\D}{r} e^{-r/\lambda_\D} + \frac{b^2}{r^2} , \label{eq:ueff}
\end{equation}
in which $+$ refers to the repulsive $(q_s q_{s^\prime} > 0)$, and $-$ to the attractive $(q_s q_{s^\prime} < 0)$ Yukawa potentials. 

\begin{figure}
\includegraphics{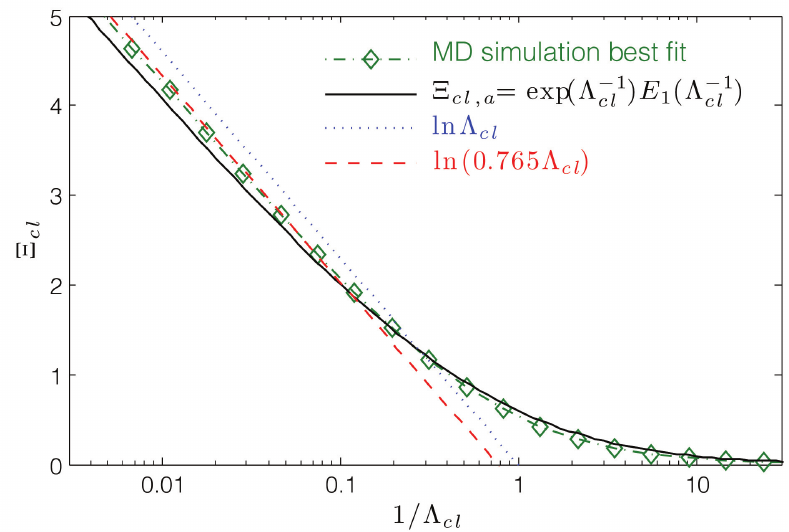}
\caption{Generalized Coulomb logarithm, $\Xi$, from the theoretical prediction of Eq.~(\ref{eq:xicl_a}) (solid line), and the best fit line to the MD simulations of Dimonte~\cite{dimo:08} (diamonds). Also shown are the weakly-coupled results from conventional (dotted line) and convergent (dashed line) kinetic theories.}
\label{fg:xid}
\end{figure}

An exact analytic solution of Eqs.~(\ref{eq:tcl})--(\ref{eq:ueff}) for $\sigma_s$ not known, but this system has been studied numerically and in asymptotic limits.\cite{maso:67,hahn:71,kilg:93,khra:02} One significant difference that arises for classical plasmas when using the screened Coulomb, rather than bare Coulomb, potential is that $\Xi$ changes for attractive or repulsive collisions. This is not a feature of the quantum regime in the first Born approximation because the solutions merge at high energy, where the Born approximation is valid. Hahn \textit{et al.}~\cite{hahn:71} have studied the smooth merging of quantum and classical solutions, which occurs when $\Lambda_{\cl} \xi^2 \simeq \Lambda_{\qm} \xi$, by accounting for the statistics of distinguishable and indistinguishable particles. 

Khrapak \textit{et al.}~\cite{khra:02} have considered $\sigma_s$ for attractive collisions between dust particles. One complication is that a potential barrier forms when $\Lambda_{\cl} \xi^2 \lesssim 1/13$. Above this critical value, a good approximation of the momentum scattering cross section is $\sigma_{s}^{\cl} \approx 4 \sigma_o \xi^{-4} \ln(1+\Lambda_{\cl}\xi^2)$.\cite{khra:02} Using this approximate cross section, Eq.~(\ref{eq:xigen}) gives
\begin{equation}
\Xi_{\cl,a} = \exp(\Lambda_{\cl}^{-1}) E_1(\Lambda_{\cl}^{-1})  \label{eq:xicl_a}
\end{equation} 
for the generalized Coulomb logarithm when $\Lambda_{\cl} \gtrsim 1/13$, which covers the weak and moderate correlation regimes. In the weakly coupled limit $\Lambda_{\cl} \gg 1$, this returns the conventional Coulomb logarithm $\Xi_{\cl,a} = \ln \Lambda_{\cl} - \gamma + \mathcal{O}(\Lambda_{\cl}^{-1})$. Figure~\ref{fg:xid} shows a favorable comparison of Eq.~(\ref{eq:xicl_a}) to the molecular dynamics simulation results of Dimonte \textit{et al.}~\cite{dimo:08}  In the simulations, $\Xi$ was determined from thermal relaxation in an electron-ion plasma: $dT_e/dt \approx 2 Q^{e-i}/3n_e$. The green line, and triangles, in Fig.~\ref{fg:xid} represent the best fit line, $\Xi= \ln(1+0.7 \Lambda_{\cl})$, of the simulation data.\cite{dimo:08} Also shown are the weakly coupled results of conventional,\cite{land:36,rose:57,lena:60} and convergent,\cite{aono:68,libo:59,brow:05,lee:84,lifs:81} kinetic theories.

\begin{figure}
\includegraphics{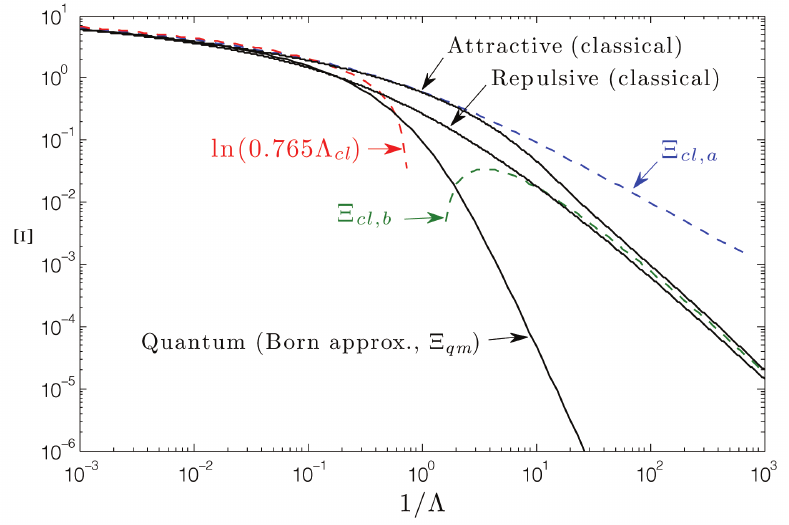}
\caption{Numerical solution of $\Xi_{\cl}(\Lambda_{\cl})$ for the attractive and repulsive Yukawa potentials, along with the asymptotic solutions from Eqs.~(\ref{eq:xicl_a}) and (\ref{eq:xicl_b}) and the quantum solution, $\Xi_{\qm}(\Lambda_{\qm})$, from Eq.~(\ref{eq:xiqm}).}
\label{fg:fg2}
\end{figure}

Beyond the barrier potential, $\Lambda_{\cl} \xi^2 \lesssim 1/13$, Khrapak \textit{et al.} show that $\sigma_s \approx \mathcal{A} \sigma_o \Lambda_{cl}^2[\ln^2(\Lambda_{\cl} \xi^2)- 2 \ln(\Lambda_{\cl}\xi^2) + \mathcal{O}(1)]$, where $\mathcal{A} \simeq 0.81$. In this strongly coupled limit, Eq.~(\ref{eq:xigen}) gives 
\begin{equation}
\Xi_{\cl,b} = (\mathcal{A}/2) \Lambda_{\cl}^2 [\ln^2(\Lambda_{\cl}) + \ln (\Lambda_{\cl}) (1-\gamma) + \mathcal{O}(1)] , \label{eq:xicl_b}
\end{equation}
for the generalized Coulomb logarithm. Note that $\mathcal{A}/2 \simeq 2/5$. The asymptotic solutions from Eq.~(\ref{eq:xicl_a}) and (\ref{eq:xicl_b}) are shown in Fig.~\ref{fg:fg2} along with direct numerical solutions of Eqs.~(\ref{eq:xigen}) and (\ref{eq:tcl})--(\ref{eq:ueff}). Both the repulsive and attractive solutions asymptote to Eqs.~(\ref{eq:xicl_a}) and (\ref{eq:xicl_b}) in the large and small $\Lambda_{\cl}$ limits, but separate in an intermediate region ($10^{-2} \lesssim \Lambda_{\cl} \lesssim 10^2$). Figure~\ref{fg:fg2} also shows the approximate quantum solution from Eq.~(\ref{eq:xiqm}). Although a given collision in any plasma may be quantum or classical, depending on $\xi$, here we assume all collisions are either quantum or classical. Accounting for both simultaneously requires $\Xi = \Xi(\Lambda_{\cl}, \Lambda_{\qm})$, which may be important in ICF plasmas that can traverse both regimes in a single implosion.  

In summary, the conventional calculation of fluid transport coefficients has been generalized to include large angle collisions, which are important in correlated plasmas. Transport is more complicated in correlated regimes, e.g., attractive and repulsive collisions can be distinguished, but the primary extension of conventional theory is a generalized Coulomb logarithm. This theory made use of the Yukawa potential for individual particles, which can often be justified regardless of the correlation strength. It is limited by the binary collision approximation, which breaks down at sufficiently strong coupling. 


The author gratefully acknowledges conversations with Dr.\ Matthew Terry, and thanks Dr.\ Will Fox, and Prof.\ Amitava Bhattacharjee for reading the manuscript. This research was supported by the US DOE Fusion Energy Postdoctoral Research Program administered by the Oak Ridge Institute for Science and Education.


\end{document}